# Proposed experiment on the continuity of quantum entanglement

Masanori Sato

*Honda Electronics Co., Ltd.,*
*20 Oyamazuka, Oiwa-cho, Toyohashi, Aichi 441-3193, Japan*

E-mail: msato@honda-el.co.jp

*Abstract* We propose experiments on quantum entanglement for investigating the Einstein Podolsky Rosen (EPR) problem with the polarization directions of photons. These experiments are performed to investigate whether the defined polarization directions in an entangled state are teleported between entangled photons. EPR-type sequential experiments are performed using a twin-photon beam and two pairs of linear polarization analyzers under the cross-Nicol condition (i.e., orthogonal to each other). If the third filter whose polarization angle is 45° is set between the first cross-Nicol filters, the beam intensity is changed from 0 to 12.5 %, and at the second cross-Nicol filters, the beam intensity is changed from 0 to 25 %. In this experiment, we predict that the "continuity of quantum entanglement" under a pure Hamiltonian evolution is detected.

1. INTRODUCTION

Bell [1] repeatedly mentioned the de Broglie-Bohm picture in his book. "Indeed it was the explicit representation of quantum nonlocality in that picture which started a new wave of investigation in this area. Let us hope that these analyses also may one day be illustrated, perhaps harshly, by some simple constructive model."

Clauser and Horne [2] and Aspect [3] proposed experiments on Bell inequalities. There were many experiments on the violation of Bell inequalities [4~8]. In 1981, Aspect et al. [6] showed experimentally the violation of Bell inequalities clearly using acousto-optical switches. Weihs et al. [8] also showed experimentally the violation of Bell inequalities by parametric down-conversion. Theories and experiments investigating Bell inequalities were summarized by Mandel and Wolf [9]. We also proposed experiments on quantum entanglement for investigating the EPR problem in the polarization direction, using the parametric down-conversion of a laser beam [10]. We showed that we can investigate whether the influence of another photon on the polarization direction [10].

We know that there exists a problem of causality, which forbids superluminal information transmission, (consequently, there is no continuity of quantum entanglement). Eberhard [11] mentioned that EPR correlation does not enable the transmission of any signal including, for example, those that are faster than light. He showed that the average spin properties of a particle are not changed by the detection of another particle which is in an entangled state. However, Bell [1] pointed out the importance of individual measurements. Holland [12] mentioned "nonlocality in the individual process, statistical locality" and also discussed the problem of signaling via quantum entanglement, "Yet the statistical compatibility of quantum mechanics with relativity in this case seems to be something of an accident". We believe that the difference between statistical and individual measurements should be tested experimentally. We think it is important to calculate it mathematically, however it is rather difficult to calculate the sequential situation of the entangled polarization direction after passing through two polarization analyzers. Therefore, there is some merit to this proposed experiment. Of course, this proposal may not be particularly new, however, at this stage, we did not find any proposal which is specially designed to test the continuity of quantum entanglement. We consider that the experiment which tests the continuity of quantum entanglement is consciously avoided owing to the violation of causality. An experimental condition that violates causality is not taken into account. For example, in the delayed-choice experiments by Hellmuth et al. [13], experimental conditions that violate causality are carefully eliminated. (Experimental conditions strictly satisfied causality.) This is the reason why there are few proposals of testing the continuity of quantum entanglement.

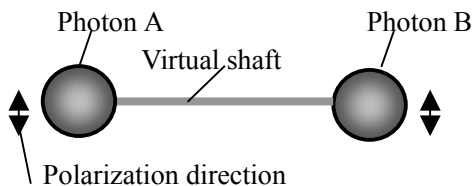

**Fig. 1** Proposed illustration of quantum entanglement: The polarization directions of photons A and B are bound by a virtual shaft. Therefore, the polarization directions of photons A and B rotate as if these photons are one particle.

We propose an experiment on the continuity of quantum entanglement using twin-photon beams and two pairs of linear polarization analyzers, and illustrations of the Bell theorem and quantum entanglement. Thereafter, we discuss causality and quantum entanglement from the viewpoints of individual measurements and statistical procedures.

## 2. PROPOSED ILLUSTRATION OF QUANTUM ENTANGLEMENT

**Figure 1** shows the proposed illustration of quantum entanglement. The polarization directions of photons A and B are bound by a virtual shaft. Therefore, the polarization directions of photons A and B rotate as if they are one particle. At this stage, the virtual shaft represents hidden variables. This model violates Bell inequalities.

## 3. PROPOSED EXPERIMENT ON CONTINUITY OF QUANTUM ENTANGLEMENT

**Figure 2** shows the experimental setup of the proposed experiment. An argon-ion laser pumps a beta-barium-borate (BBO) crystal to produce polarization entangled photon pairs via spontaneous parametric down-conversion [8]. In this experiment, we detect the intensity of twin-photon beams.

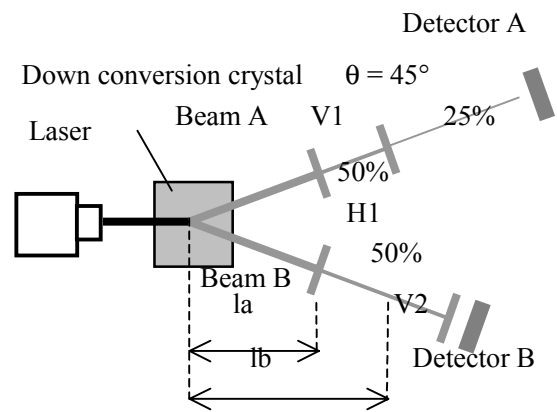

**Fig. 2** Experimental setup:
A down-conversion crystal generates a twin-photon beam by the type-2 down conversion of an incident laser beam. Two polarization analyzers, H1 and V2, are set orthogonal to each other, that is, under the cross-Nicol condition. If a polarization analyzer, whose polarization angle is 45°, θ= 45°, is set after the polarization analyzer V1, we can test whether the condition of the polarization direction of beam A is teleported to beam B.

We do not check the correlation of entangled photon pairs. Therefore, the proposed experiment is feasible. A down-conversion crystal generates a twin-photon beam by the type-2 down-conversion of an incident laser beam. Beam A is filtered by a vertical polarization analyzer, V1, and thereafter by a polarization analyzer whose polarization angle is 45°, θ = 45°. The photon intensities of beam A are changed from 100% to 50% by passing through the polarization analyzer, V1, and from 50% to 25% by passing through the polarization analyzer θ = 45°. Beam B is filtered by a horizontal polarization analyzer, H1, and thereafter by a vertical polarization analyzer V2. In this experiment, if beam A does not pass through the polarization analyzer θ = 45°, beam B is not detected by detector B (i.e., the photon intensity detected by detector B is 0). This is because the horizontal polarization analyzer H1 is set orthogonal to the vertical polarization analyzer V2 (i.e., cross-Nicol condition).

In the experimental setup shown in **Fig. 2**, using the polarization analyzers V1 and θ = 45°, if entanglement continues after the beams pass through the polarization analyzers V1 and H1, there is a possibility that we can detect photons on detector B. We assume that quantum entanglement in the polarization direction occurs as shown in **Fig. 1**, i.e., the polarization directions are bound by virtual shaft.

In this experiment, we can predict the following two possible answers: (1) The first polarizer projects the entangled state on a nonentangled one, or (2) the continuity of entanglement under a pure Hamiltonian evolution is detected. At this stage, when the experiments have not been carried out, we would consider the latter.

However, the continuity of entanglement indicates that we can transmit a signal through quantum entanglement. For example, the detection by detector B provides information on using the polarization analyzer θ = 45°, because we cannot detect photons using detector B, when the polarization analyzers H1 and V2 are set orthogonal to each other.

Beams A and B are in an entangled state, and if photons A and B are in an entangled state after passing through the polarization analyzers V1 and H1 at time $t_1$, thereafter, at time $t_2$ when photon A passes through the polarization analyzer θ = 45°, photon B will change its polarization direction. **Figure 3** is an illustration, which shows the entanglement of polarization direction. This indicates that the polarization direction of photon B is not orthogonal to the polarization analyzer V2; therefore, photon B can be detected by detector B with a 50%

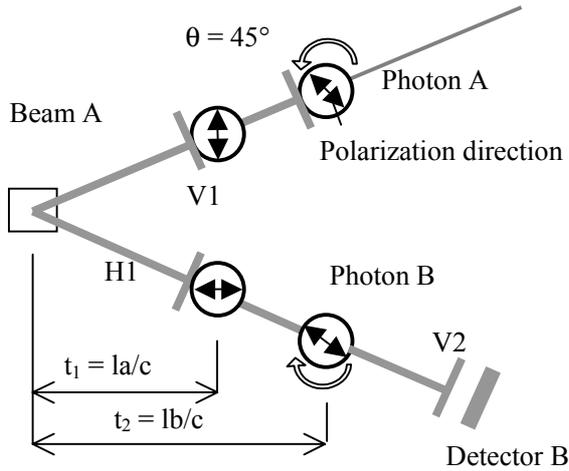

**Fig. 3** Proposal of the rotation of polarization direction through quantum entanglement: The illustration shows the entanglement of polarization directions. The polarization direction of photon B is not orthogonal to the polarization analyzer V2; therefore, photon B can be detected by detector B with a 50% probability. We can detect a beam intensity of 25% by detector B

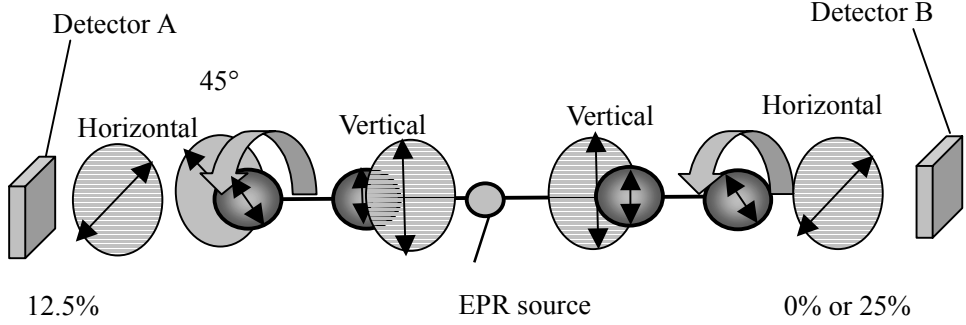

Fig. 4  Illustration of the rotation of polarization direction: A photon pair from the EPR source is filtered by a vertical filter pair. Then photon A is filtered by a 45° filter. We predict that information on the polarization direction is teleported to photon B, i.e., the polarization direction of photon B is changed. Thereafter, photons A and B are filtered by a horizontal filter pair. We obtain a beam intensity of 12.5% for photon A at detector A. At detector B, we would predict a detection of beam intensity of 25% rather than 0%. The vertical and horizontal filters are set orthogonal to each other (i.e., cross-Nicol condition), then nonentangled photons cannot pass through the filter pair. Therefore, we can carry out the experiments with little disturbance of nonentangled photons.

probability. We can detect a beam intensity of 25% by detector B.

**Figure 4** shows the conceptual illustration of total spin 0 state. Type-1 down conversion is used. A photon pair from the EPR photon source is filtered by a vertical filter pair. Then photon A is filtered by a 45° filter. Thereafter, photons A and B are filtered by a horizontal filter pair. We obtain a beam intensity of 12.5% for photon A using detector A. At detector B, we would predict a detection of beam intensity of 25% rather than 0%. This indicates that the condition of the polarization direction of photon B is teleported to that of photon A. However, in this experimental setup, the existence of the 45° filter can be a superluminal signal (see discussion).

In **Fig. 4** the vertical and horizontal filters are set orthogonal to each other (i.e., cross-Nicol condition), then nonentangled photons cannot pass through the filter pair and not be detected by detector B. We can carry out the experiments with little disturbance of nonentangled photons.

## 3. DISCUSSION
### A. Quantum entanglement and interaction

We define quantum entanglement and interaction clearly. Quantum entanglement is permanent relationship, for example twin-photon generated by down-conversion will be in an entangled state permanently. The origin of entangled photons is the down-conversion of the photon energy hν. Quantum entanglement continues forever (i.e., conservation of total spin 0 is satisfied if observed at any time.)

However, interaction is temporal relationship, for example collision of two particles. The two particles exchange momentum, spin, and energy satisfying conservation law. The meaning of "interaction is temporal" is "interaction occurs in a finite period," and does not continue. The correlation of spin is not permanent, it will be changed by another interaction.

Bohm's representation of quantum entanglement related to the loss of nonlocality in the classical limit provides a very clear expression [15]. According to his writings and the expression of the equations, we can illustrate Bohm's picture. Bohm clearly indicated in his book through the expression of wave functions that in the interaction of atom A (where atom A and atom B are in an entangled state) with atom C, the spin conservation not only between A and C but also between A and B is satisfied. Therefore, **Fig. 5** can be drawn. The conservation of the spin between atoms A with C is represented as i + j = i' + j', where i and j are the spins and ' indicates the state after interaction. The spin of atom A is changed from j to j', and simultaneously, the spin of atom B is changed from -j to -j'. This indicates that a total spin 0 between atoms A and B is satisfied. This is entanglement, and is clearly different from interaction.

According to Bohm, the origin of the change in the spin of atom B is the quantum potential. **Figure 5** appears to indicate that quantum entanglement continues after the interaction occurs.

In **Fig. 4**, photons A and B are in an entangled sate however, they cannot be in an entangled state with the electrons in the polarization analyzers. Photons A and B can only interact with the electrons in the polarization analyzers. The entanglement of photons A and B continues after passing the polarization analyzers.

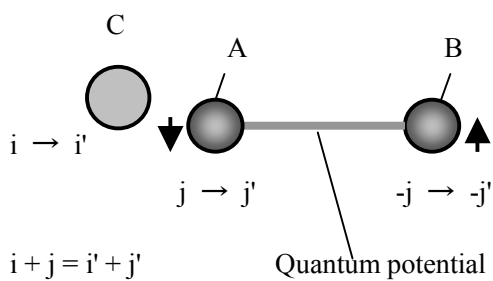

Fig. 5 Conceptual illustration of interaction according to Bohm's writings and the expression of equations. Atoms A and B are entangled, atom C interact with A and B, spin i and j interact, however Atoms A and C cannot be in an entangled state. Atoms A and B are still in an entangled state, i.e., entanglement continues.

B. Suspend restriction of causality

Tentatively, we restrict causality to mean information transmission through photons, therefore, superluminal information transmissions through photons are not possible, because information is transmitted by photons. However, information transmission, which does not occur through photons would not be restricted by causality. Quantum entanglement does not occur through photons, therefore, there is a possibility that information transmission is not restricted by light speed. Of course, information transmission through quantum entanglement should be tested experimentally.

C. Individual measurement and statistical procedures

Bell pointed out the importance of individual measurements, as evidenced by his statement, "The de Broglie-Bohm picture agrees with quantum mechanics in having the eigenvalues as the result of individual measurements." (Bell [1]). However, there have been only few reports on individual measurements. Holland pointed out that "the quantum theory of motion permits more detailed prediction to be made pertaining to the individual process." (Holland [12]).

On the other hand, there are many reports arguing against superluminal information transmission through quantum entanglement, for example, those by Eberhard [11], Ghirardi et al. [14], and Holland [12]. They pointed out that information disappears after averaging. Causality is satisfied after averaging i.e., information disappears after averaging. For example, we cannot distinguish the signals between 0101 and 1100 after averaging. We can only obtain numerical value of 0.5, which is the average of four numerical values. We believe that information disappears after averaging. However, it does not exclude the possibility of information transmission via individual measurements.

## 4. CONCLUSIONS

We examined the proposed experiments on the continuity of quantum entanglement from the viewpoints of causality and individual measurements. In these experiments, we detect the intensity of twin-photon beams, but not the correlation of twin-photon. Therefore, it is easy to perform these experiments. The merit of these experiments in practice is the possibility of new interesting physics. At this stage, when the experiments have not yet been carried out, we consider the existence of the continuity of quantum entanglement.

REFERENCES
1) J. S. Bell, "*Speakable and unspeakable in quantum mechanics*" (Cambridge University Press, Cambridge, 1987).
2) J. Clauser and M. A. Horne, "Experimental consequences of objective local theories," Phys. Rev. D, **10**, 526 (1974).
3) A. Aspect, "Proposed experiment to test the nonseparability of quantum mechanics," Phys. Rev. D, **14**, 1944 (1976).
4) J. F. Clauser, "Experimental Investigation of a Polarization Correlation Anomaly," Phys. Rev. Lett. **36**, 1223, (1976).
5) S. J. Freedman and J. F. Clauser, "Experimental Test of Local Hidden-Variable Theories," Phys. Rev. Lett. **28**, 938, (1972).
6) A. Aspect, P. Grangier and G. Roger, "Experimental Tests of Realistic Local Theories via Bell's Theorem," Phys. Rev. Lett., **47**, 460 (1981).
7) Z. Y. Ou and L. Mandel, "Violation of Bell's Inequality and Classical Probability in a Two-Photon Correlation Experiment," Phys. Rev. Lett. **61**, 50, (1988).
8) G. Weihs, T. Jennewein, Ch. Simon, H. Weinfurter, and A. Zeilinger, "Violation of Bell's inequality under strict Einstein locality conditions," Phys. Rev. Lett., **81**, 5039 (1998).
9) L. Mandel and E. Wolf, "*Optical coherence and quantum optics*" (Cambridge University Press, Cambridge, 1995).
10) M. Sato, "Quantum entanglement and signaling," The 10$^{th}$ JST International Symposium Proceedings (Tokyo, Japan, March 2002).
11) P. H. Eberhard, "Bell's Theorem and Different Concepts of Locality," Nuovo Cimento, **46B**, 392 (1978).
12) P. R. Holland, "*The Quantum Theory of Motion*" (Cambridge University Press, Cambridge, 1994).
13) T. Hellmuth, H. Walther, A. Zajonc, and W. Schleich, "Delayed-choice experiments in quantum interference," Phys. Rev. A, **35**, 2532 (1987).
14) G. C. Ghirardi, A. Rimini, and T. Weber, "A general argument against superluminal transmission through the quantum mechanical measurement process," Lett. Nuovo Cimento, **27**, 293 (1980).
15) D. Bohm and B. J. Hiley, "*The Undivided Universe*, " (Routledge, London, 1993).